# Ginzburg-Landau analysis of the critical temperature and the upper critical field for three-band superconductors

Y. Yerin[1], S.-L. Drechsler[2], G. Fuchs[2]

[1] *B. Verkin Institute for Low Temperature Physics and Engineering of the National Academy of Sciences of Ukraine, 47 Lenin Ave. Kharkov 61103, Ukraine*

[2] *Leibniz-Institute of Solid State Research and Material Science (IFW-Dresden), D-01171 Dresden, Germany*

e-mail: yerin@ilt.kharkov.ua

ABSTRACT

Using the microscopic formalism of Eilenberger equations, a three-band Ginzburg-Landau theory for the intraband dirty limit and clean interband scattering case is derived. Within the framework of this three-band Ginzburg-Landau theory, expressions for the critical temperature $T_c$ and the temperature dependence of the upper critical field $H_{c2}$ are obtained. Based on some special cases of the matrix of interaction constants, we demonstrate the influence of the sign of the interband interaction on the critical temperature and the upper critical field as compared with a two-band superconductor where it plays no role. We study also analytically and numerically the effect of its magnitude.

## 1. Introduction

Most of the real superconductors exhibit multiple Fermi surface sheets. Especially in the context of recently discovered iron pnictide superconductors [1] it becomes more and more clear that the frequently adopted two-band approach doesn't allow for quantitative fits for various physical properties and a study of more complex effective three-band [2,3] or even higher multiple band cases is necessary. According to the recently introduced generalized density functional theory including also an superconducting order parameter within the framework of the Bogolyubov - de Gennes theory, molecular hydrogen at significantly high pressure, demonstrates a Fermi surface with different and disconnected sheets, whose electrons are strongly coupled with inter- and intramolecular phonon modes [4]. This combination gives rise to anisotropic three-band superconductivity with a critical temperature up to room temperature.

Other examples of superconductors, where a three-band approach should be used for the description of superconducting properties, are doped fullerides. In Refs. 5, 6 it was pointed out that an important factor determining the magnitudes of the critical temperature of the superconducting doped fullerenes is the extent to which the Cooper pairs are delocalized over the three bands at the Fermi level.
In the context of exotic three-band superconductors also the first *p*-wave superconductor $Sr_2RuO_4$ is worth to be mentioned [7]. Hence, the study of three-band superconductors is not an academic problem but a challenge to study in more detail the mentioned above complex real systems. For that purpose the present approach provides a reasonable starting point.

Finally, within a phenomenological approach (extended Ginzburg-Landau approach taking into account higher order terms) for three-band superconductors under certain conditions novel stable topological defects like phase solitons and unusual fractional vortices have been predicted [8-10]. Moreover, recently it was found that multi-band superconductivity with weak interband coupling may exhibit a hidden critical point [11]. To understand which of the real compounds will meet these special conditions requires a comprehensive description within these phenomenological models in order to detect the predicted and mentioned above peculiarities experimentally.

In spite of the natural observations of three gaps, less is known for other thermodynamic properties. In this context the interpretation of experimental upper critical field data in terms of multiband models beyond single-band strong coupling theories [12] and two-band model approximations [13-23] is highly desirable. From a theoretical point of view, the possibility of an unusual broken time-reversal symmetry and accompanying frustration phenomena for the ground states of systems with odd-numbered bands and repulsive interband couplings between them has been attracted considerable attention [24-29].

Here, we derive Ginzburg-Landau (GL) equations for three-band superconductors from quasi-classical Usadel equations for the case of a dirty superconductor in the sense of strong intraband scattering by non-magnetic impurities. However, at the same we ignore for the sake of simplicity impurity induced elastic interband scattering. Its effect is qualitatively well-known: a reduction of the critical temperature, especially in the case of repulsive interband couplings and a corresponding change of the symmetry of the ground state towards a standard so-called $s_{++}$ -symmetry provided the intraband couplings are strong enough to yield a finite $T_c$-value. In cases when the different bands involve different orbitals that scattering can be weak and ignored in the first approximation. Anyhow, in principle, this scattering can be also incorporated into a Ginzburg-Landau functional as has been shown for the case of two-band superconductivity for instance in Refs. 30, 31. We postpone the consideration of this interesting and important issue for the general three-band situation to a future study. Finally, we note that our theory in the present form cannot be applied also to cases with nodal order parameters as in the *d*- or *p*-wave cases since there nonmagnetic intraband impurities are pair-breaking like magnetic impurities in conventional $s_{++}$-superconductors.

The aim of the present paper is twofold: (i) to provide general equations to be applied in forthcoming papers to real materials with the aim to find real candidates among them for the experimental detection of the predicted exotic properties mentioned above and (ii) to consider some special cases which demonstrate clearly the richness of higher order multiband models as compared to frequently used two-band cases. Thus, it is not the aim of our paper to describe the new and subtle physics related to unusual vortices and other exotic excitations mentioned above, the more that there might be limitations for such problems to be attacked within a simple GL-approach [8-10] as we use here. Instead our results for unusual shapes of the upper critical field $H_{c2}(T)$ reported below might be helpful to select possible promising candidates among the increasing number of real materials suitable for such searches. (iii) To find preliminary parameter and temperature regions, although formally beyond the formal validity of a Ginzburg-Landau theory based description, where the unusual behavior obtained here (e.g. low-temperature peculiarities of the upper critical field, see below) suggests to perform calculations also within more sophisticated approaches to check or to refine our findings but with much higher numerical efforts.

## 2. Derivation of Ginzburg-Landau equations

Generally, Usadel equations can be derived from the Eilenberger equations using the same formalism as for a single-band (see for instance Refs. 32, 33) or a two-band superconductor (see A. Gurevich [16]). In the present three-band case, the Usadel equations take the following form:

$$\omega f_1 - \frac{D_1}{2}\left(g_1\Pi^2 f_1 - f_1\Pi^2 g_1\right) = \Delta_1 g_1 + \Gamma_{12}\left(g_1 f_2 - g_2 f_1\right) + \Gamma_{13}\left(g_1 f_3 - g_3 f_1\right), \quad (1)$$

$$\omega f_2 - \frac{D_2}{2}\left(g_2\Pi^2 f_2 - f_2\Pi^2 g_2\right) = \Delta_2 g_2 + \Gamma_{21}\left(g_2 f_1 - g_1 f_2\right) + \Gamma_{23}\left(g_2 f_3 - g_3 f_2\right), \quad (2)$$

$$\omega f_3 - \frac{D_3}{2}\left(g_3\Pi^2 f_3 - f_3\Pi^2 g_3\right) = \Delta_3 g_3 + \Gamma_{31}\left(g_3 f_1 - g_1 f_3\right) + \Gamma_{32}\left(g_3 f_2 - g_2 f_3\right). \quad (3)$$

These Usadel equations (1-3) have to be supplemented with three self-consistency equations for the three order parameters $\Delta_i$:

$$\Delta_i = 2\pi T \sum_j \sum_{\omega>0}^{\omega_D} \lambda_{ij} f_j \, . \quad (4)$$

Here $\Pi \equiv \nabla + \frac{2\pi i}{\Phi_0}\boldsymbol{A}$. The index $i$ =1-3 in Eq. (4) denotes the band number. The Green's functions $g_i$ and $f_i$ are connected by the normalization condition $g_i^2 + |f_i|^2 = 1$ and depend on the spatial coordinates and the Matsubara frequencies $\omega = (2n+1)\pi T$. $D_i$ are the intraband diffusivities due to nonmagnetic impurity scattering, $N_i$ are the partial density of states on the Fermi surface for the electrons of the *i*-th band, $\lambda_{ij}$ are the dimensionless interaction constants (electron-phonon (boson), electron-electron, etc. couplings depending on the pairing mechanism) and $\Gamma_{ij}$ are the interband scattering rates, which take into account the effect of non-magnetic impurity scattering.

Neglecting the interband (impurity induced) scattering terms and using the method of successive approximations, we obtain the corresponding GL-equations, valid strictly speaking, in the vicinity of $T_c$ (see Appendix A):

$$\begin{cases} \left[\frac{\pi D_1}{8T_c}\Pi^2\Delta_1 - \frac{7\zeta(3)}{8\pi^2T_c^2}\left|\Delta_1\right|^2\Delta_1 + \Delta_1\ln\left(\frac{2\gamma\langle\omega_0\rangle}{\pi T}\right)\right]N_1\det(\lambda) = \\ \Delta_1(\lambda_{22}\lambda_{33}-\lambda_{23}\lambda_{32})N_1 - \Delta_2(\lambda_{12}\lambda_{33}-\lambda_{13}\lambda_{32})N_2 + \Delta_3(\lambda_{12}\lambda_{23}-\lambda_{13}\lambda_{22})N_3, \\ \left[\frac{\pi D_2}{8T_c}\Pi^2\Delta_2 - \frac{7\zeta(3)}{8\pi^2T_c^2}\left|\Delta_2\right|^2\Delta_2 + \Delta_2\ln\left(\frac{2\gamma\langle\omega_0\rangle}{\pi T}\right)\right]N_2\det(\lambda) = \\ = -\Delta_1(\lambda_{21}\lambda_{33}-\lambda_{23}\lambda_{31})N_1 + \Delta_2(\lambda_{11}\lambda_{33}-\lambda_{13}\lambda_{31})N_2 - \Delta_3(\lambda_{11}\lambda_{23}-\lambda_{21}\lambda_{13})N_3, \\ \left[\frac{\pi D_3}{8T_c}\Pi^2\Delta_3 - \frac{7\zeta(3)}{8\pi^2T_c^2}\left|\Delta_3\right|^2\Delta_3 + \Delta_3\ln\left(\frac{2\gamma\langle\omega_0\rangle}{\pi T}\right)\right]N_3\det(\lambda) = \\ = \Delta_1(\lambda_{21}\lambda_{32}-\lambda_{22}\lambda_{31})N_1 - \Delta_2(\lambda_{11}\lambda_{32}-\lambda_{31}\lambda_{12})N_2 + \Delta_3(\lambda_{11}\lambda_{22}-\lambda_{12}\lambda_{21})N_3. \end{cases} \tag{5}$$

Let's introduce

$$\alpha_1 = \left(\ln\left(\frac{2\gamma\langle\omega_0\rangle}{\pi T}\right) - \frac{\lambda_{22}\lambda_{33}-\lambda_{23}\lambda_{32}}{\det(\lambda)}\right)N_1 = (l-a_1)N_1,$$

$$\alpha_2 = \left(\ln\left(\frac{2\gamma\langle\omega_0\rangle}{\pi T}\right) - \frac{\lambda_{11}\lambda_{33}-\lambda_{13}\lambda_{31}}{\det(\lambda)}\right)N_2 = (l-a_2)N_2,$$ and

$$\alpha_3 = \left(\ln\left(\frac{2\gamma\langle\omega_0\rangle}{\pi T}\right) - \frac{\lambda_{11}\lambda_{22}-\lambda_{12}\lambda_{21}}{\det(\lambda)}\right)N_3 = (l-a_3)N_3,$$

where $l = \ln\left(\frac{2\gamma\langle\omega_0\rangle}{\pi T}\right)$, $a_1 = \frac{\lambda_{22}\lambda_{33}-\lambda_{23}\lambda_{32}}{\det(\lambda)}$, $a_2 = \frac{\lambda_{11}\lambda_{33}-\lambda_{13}\lambda_{31}}{\det(\lambda)}$, $a_3 = \frac{\lambda_{11}\lambda_{22}-\lambda_{12}\lambda_{21}}{\det(\lambda)}$ ,

and $\det(\lambda)$ is the determinant of the matrix $\lambda_{ij}$. Note that $a_i$ are the minors $M_{ij}$ of the matrix of interaction constants lying on the main diagonal, i.e. $a_i = M_{ii}/\det(\lambda)$.

Furthermore, we denote the effective interband interaction coefficients as:

$$\gamma_{12} = \frac{(\lambda_{12}\lambda_{33}-\lambda_{13}\lambda_{32})N_2}{\det(\lambda)} = \tilde{\gamma}_{12}N_2, \qquad \gamma_{13} = \frac{(\lambda_{13}\lambda_{22}-\lambda_{12}\lambda_{23})N_3}{\det(\lambda)} = \tilde{\gamma}_{13}N_3,$$

$$\gamma_{21} = \frac{(\lambda_{21}\lambda_{33}-\lambda_{23}\lambda_{31})N_1}{\det(\lambda)} = \tilde{\gamma}_{21}N_1, \qquad \gamma_{23} = \frac{(\lambda_{11}\lambda_{23}-\lambda_{13}\lambda_{21})N_3}{\det(\lambda)} = \tilde{\gamma}_{23}N_3,$$

$\gamma_{31} = \frac{(\lambda_{31}\lambda_{22} - \lambda_{21}\lambda_{32})N_1}{\det(\lambda)} = \tilde{\gamma}_{31}N_1$, and $\gamma_{32} = \frac{(\lambda_{11}\lambda_{32} - \lambda_{31}\lambda_{12})N_2}{\det(\lambda)} = \tilde{\gamma}_{32}N_2$, where $\tilde{\gamma}_{ij} = (-1)^{i+j+1} M_{ji} / \det(\lambda)$.

Then, the GL-equations using also $K_i = \frac{\pi D_i N_i}{8T_c}$ and $\beta_i = \frac{7\zeta(3)N_i}{8\pi^2 T_c^2}$, can be rewritten finally as:

$$\begin{cases} K_1\Pi^2\Delta_1 - \beta_1|\Delta_1|^2\Delta_1 + \alpha_1\Delta_1 + \gamma_{12}\Delta_2 + \gamma_{13}\Delta_3 = 0, \\ K_2\Pi^2\Delta_2 - \beta_2|\Delta_2|^2\Delta_2 + \alpha_2\Delta_2 + \gamma_{21}\Delta_1 + \gamma_{23}\Delta_3 = 0, \\ K_3\Pi^2\Delta_3 - \beta_3|\Delta_3|^2\Delta_3 + \alpha_3\Delta_3 + \gamma_{31}\Delta_1 + \gamma_{32}\Delta_2 = 0. \end{cases} \quad (6)$$

# 3. The critical temperature of a three-band superconductor

It can be shown (see Appendix B) that the linearized system of Eqs. (6) for the determination of $T_c$ is equivalent to the secular equation for the coupling matrix:

$$\begin{pmatrix} \lambda_{11} - \Lambda & \lambda_{12} & \lambda_{13} \\ \lambda_{21} & \lambda_{22} - \Lambda & \lambda_{23} \\ \lambda_{31} & \lambda_{32} & \lambda_{33} - \Lambda \end{pmatrix} = 0\,. \quad (7)$$

The critical temperature is given by the general expression (in units with $k_B = 1$ and $\hbar = 1$):

$$T_c = \frac{2\gamma\langle\omega_0\rangle}{\pi}\exp\left(-\frac{1}{\Lambda^{(r)}}\right), \quad (8)$$

where $\Lambda^{(r)}$ is the largest positive real eigenvalue of the matrix $\lambda_{ij}$ and $\langle\omega_0\rangle$ is the cut-off frequency in the spirit of a BCS-type approach. Within a more microscopical based strong coupling (Eliashberg-theory) picture $\langle\omega_0\rangle$ represents an effective frequency, which reflects the energy of the involved bosons which provide the glue for the superconducting pairing. Thereby it's assumed for the sake of simplicity that this energy is roughly the same for all interaction channels. Notice that in accordance with the Anderson-theorem, the intraband impurity scatterings measured by the diffusivities $D_i$ have been dropped out.

Here, our main interest is focused on the influence of the signs of the interband couplings constant.

If the third band is absent or decoupled from the first two bands, Eq. (7) naturally reduces to the well-known case for two-band superconductivity:

$$[\Lambda - \lambda_{33}]\left[\Lambda^2 - (\lambda_{11} + \lambda_{22})\Lambda + (\lambda_{11}\lambda_{22} - \lambda_{12}\lambda_{21})\right] = 0\,, \tag{9}$$

and the corresponding solution reads:

$$\Lambda_{1,2}^{(\text{2-band})} = \frac{\lambda_{11} + \lambda_{22} \pm \sqrt{(\lambda_{11} - \lambda_{22})^2 + 4\lambda_{12}\lambda_{21}}}{2}\,,\ \lambda_3 = \lambda_{33}\,. \tag{10}$$

Let's introduce $\lambda_+ = \lambda_{11} + \lambda_{22} + \lambda_{33}$, $m = M_{11} + M_{22} + M_{33}$ and $w = \det(\lambda)$. Using these notations we rewrite Eq. (7) as:

$$\Lambda^3 - \lambda_+\Lambda^2 + m\Lambda - w = 0\,. \tag{11}$$

The roots of Eq. (11) are

$$\Lambda_1 = \frac{\lambda_+}{3} + \frac{1}{3}\sqrt[3]{\frac{1}{2}\left(2\lambda_+^3 - 9\lambda_+ m + 27w\right) + \sqrt{Q}} + \frac{1}{3}\sqrt[3]{\frac{1}{2}\left(2\lambda_+^3 - 9\lambda_+ m + 27w\right) - \sqrt{Q}}\,, \tag{12}$$

$$\Lambda_2 = \frac{\lambda_+}{3} - \frac{1 - i\sqrt{3}}{6}\sqrt[3]{\frac{1}{2}\left(2\lambda_+^3 - 9\lambda_+ m + 27w\right) + \sqrt{Q}} - \frac{1 + i\sqrt{3}}{6}\sqrt[3]{\frac{1}{2}\left(2\lambda_+^3 - 9\lambda_+ m + 27w\right) - \sqrt{Q}}\,, \tag{13}$$

$$\Lambda_3 = \frac{\lambda_+}{3} - \frac{1 + i\sqrt{3}}{6}\sqrt[3]{\frac{1}{2}\left(2\lambda_+^3 - 9\lambda_+ m + 27w\right) + \sqrt{Q}} - \frac{1 - i\sqrt{3}}{6}\sqrt[3]{\frac{1}{2}\left(2\lambda_+^3 - 9\lambda_+ m + 27w\right) - \sqrt{Q}}\,, \tag{14}$$

where the discriminant $Q$ given by the expression

$Q = \dfrac{4\lambda_+^3 w - \lambda_+^2 m^2 - 18\lambda_+ m w + 27 w^2 + 4m^3}{108}$ has been used.

Depending on the sign of the discriminant $Q$, we have three distinct real roots, if $Q < 0$, one real root and two complex conjugated roots, if $Q > 0$ and multiple real roots, if $Q = 0$. Below we will study some simple special cases.

**(i)** First we readdress the case considered in Ref. 24 with equal intraband and interband couplings., i.e. $\lambda_{11} = \lambda_{22} = \lambda_{33} = k_0$ and $\lambda_{12} = \lambda_{21} = \lambda_{13} = \lambda_{31} = \lambda_{23} = \lambda_{32} = k_1$, where $k_0, k_1 > 0$. It means $\lambda_+ = 3k_0$, $m = 3k_0^2 - 3k_1^2$ and $w = k_0^3 + 2k_1^3 - 3k_0 k_1^2$. Substituting these redefined parameters into the expression for the discriminant $Q$ we obtain that $Q = 0$ . Then, Eq. (11) has two real roots for all $k_0$ and $k_1$:

$$\begin{aligned} \Lambda_1 &= k_0 + 2k_1, \\ \Lambda_{2,3} &= k_0 - k_1. \end{aligned} \tag{15}$$

As mentioned above, we must choose the largest eigenvalue, i.e. $\Lambda_1 = k_0 + 2k_1$ for $k_1 > 0$. Compared with two-band superconductivity for which in our terms $\Lambda^{(\text{2-band})} = k_0 + |k_1|$ holds, the presence of the third band *enhances* the critical temperature (according to Eqs. (8) and (10)).

In case of repulsive interband couplings one has $k_1 < 0$, which for two-band superconductivity leads to the so-called $s_\pm$-pairing symmetry frequently discussed in the context of iron pnictides. In other words, here we are left with $\lambda_{11} = \lambda_{22} = \lambda_{33} = k_0$ and $\lambda_{12} = \lambda_{21} = \lambda_{13} = \lambda_{31} = \lambda_{23} = \lambda_{32} = -|k_1|$. Then, the largest eigenvalue is $\Lambda_{2,3} = k_0 + |k_1|$ and it's the same as for the two-band superconductor, i.e. for repulsive interband couplings the inclusion of a third band doesn't affect $T_c$ in contrast to the case of attractive interband couplings.

This is a noteworthy qualitative feature of a three-band superconducting system that distinguishes it from a two-band superconductor where the sign of the interband coupling constant plays no role (see Eq. (9)). So we can conclude that at least for this set of parameters the existence of the third band increases the critical temperature for attractive interband couplings or leaves it unchanged for repulsive counter parts.

**(ii)** Let us consider another case of a non-trivial three-band superconductor which was investigated in Refs. 26, where the coupling matrix with repulsive interband interaction constants, only, has the form:

$$\begin{pmatrix} 0 & k_1 & k_1 \\ k_1 & 0 & k_2 \\ k_1 & k_2 & 0 \end{pmatrix}. \tag{16}$$

From this matrix we have for the discriminant of Eq. (11) $Q = -\frac{\left(k_1^2 - k_2^2\right)^2 \left(8k_1^2 + k_2^2\right)}{27}$, which is non-positive for all $k_1$ and $k_2$. Hence, Eq. (11) has three real solutions as in the previous case:

$$\Lambda_1 = -k_2, \tag{17}$$

$$\Lambda_2 = \frac{1}{2}k_2 - \frac{1}{2}\sqrt{8k_1^2 + k_2^2}, \tag{18}$$

$$\Lambda_3 = \frac{1}{2}k_2 + \frac{1}{2}\sqrt{8k_1^2 + k_2^2}. \tag{19}$$

Next, we determine the regions of $k_1$ and $k_2$, where the first root yields the largest eigenvalue, then the region for the second root and finally that for the third one, respectively.

We plot the peculiar phase diagram (Fig. 1), which demonstrates the distribution of eigenvalues versus the values of interband coefficients and found out that for arbitrary real non-zero values of $k_1$ and $k_2$ $\Lambda_1$ and $\Lambda_3$ yield always the largest eigenvalue (in the corresponding regions) for this matrix of the interaction constants.

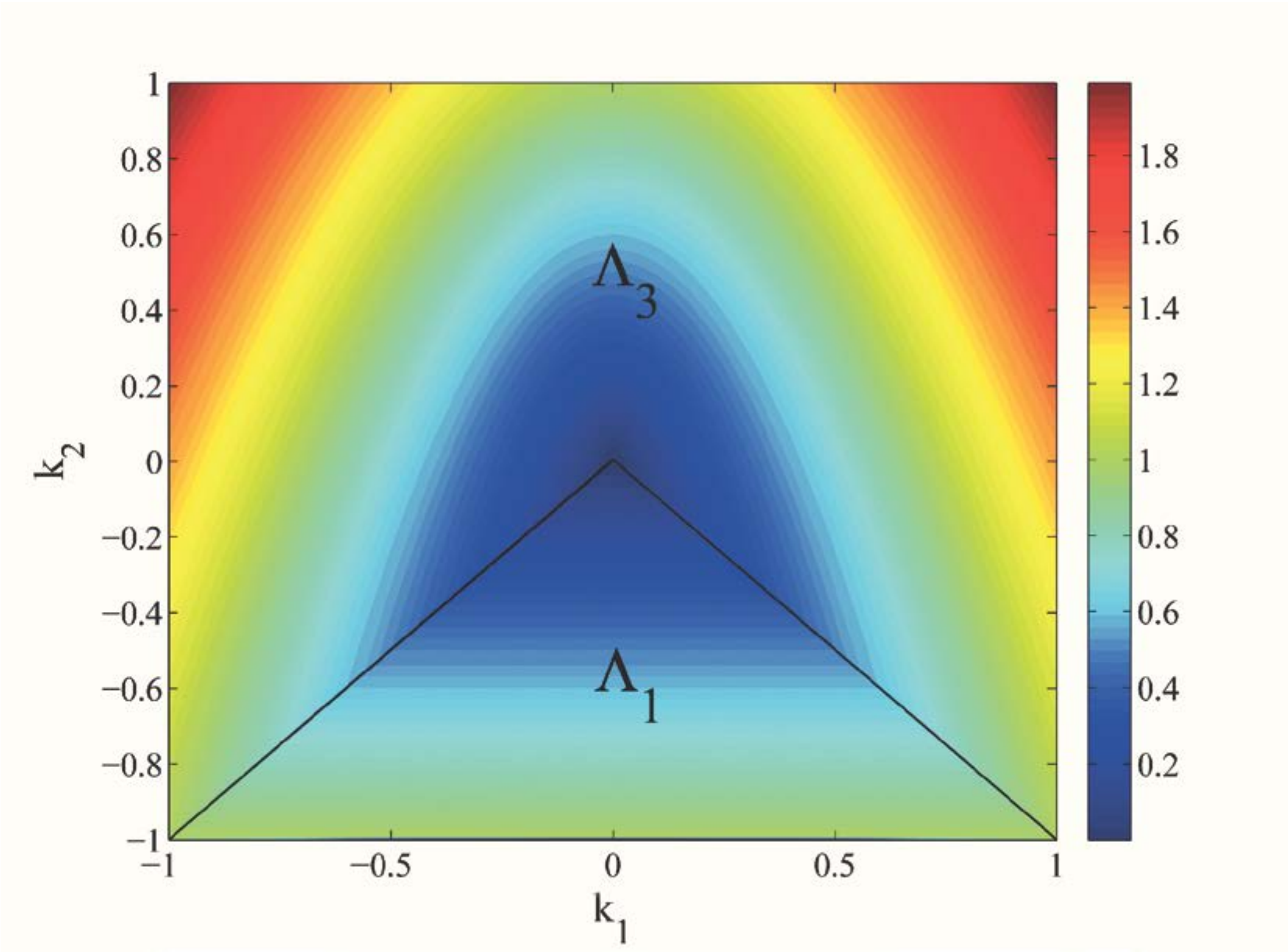

Fig. 1 (Color). The distribution of the largest eigenvalues of the matrix of the interaction constants of a three-band superconductor with interband couplings, only (see the matrix (16)). The black lines divide the figure on two regions with corresponding eigenvalues. The dark red and the dark blue parts exhibit the highest and the lowest $T_c$-value, respectively.

The corresponding two-band superconductor has an eigenvalue of $\Lambda^{(2\text{-band})} = |k_1|$ and the presence of the third band always leads to an enhancement of $T_{\text{c}}$. Thereby the enhancement for attractive couplings exceeds that for repulsive ones for the same modulo $|k_2|$.

**(iii)** The experimental data for some iron-based pnictide superconductors have been described in the literature [2, 3] in terms of a reduced three-band model with the matrix of interaction constants:

$$\begin{pmatrix} \lambda_{11} & \lambda_{12} & \lambda_{13} \\ \lambda_{21} & \lambda_{22} & 0 \\ \lambda_{31} & 0 & \lambda_{33} \end{pmatrix}, \qquad (20)$$

where band 1 is a hole band centered around the Γ-point and band 3 is an electron band centered at the corner of the Brillouin zone to be connected by the nesting vector with band 1. Band 2 was attributed to another electron or hole band in the case of the electron (Co) doped Ba-122 system [2] and the strongly hole (Na) doped Ca-122 system [3], respectively. For all these cases the results obtained in the present paper might be of potential interest for the description of magnetic properties of such iron-based superconductors to be considered elsewhere.
For this matrix the secular equation (11) reads:

$$-\lambda_{31}\lambda_{13}\left(\lambda_{22}-\Lambda\right)+\left(\lambda_{33}-\Lambda\right)\left[\left(\lambda_{11}-\Lambda\right)\left(\lambda_{22}-\Lambda\right)-\lambda_{12}\lambda_{21}\right]=0 . \quad (21)$$

If we assume that the intraband interactions for the second and the third band coincide, then Eq. (21) reduces to

$$\left(\lambda_{11}-\Lambda\right)\left(\lambda_{22}-\Lambda\right)-\lambda_{12}\lambda_{21}-\lambda_{31}\lambda_{13}=0 , \quad (22)$$

with the solutions

$$\Lambda_{1,2}=\frac{\lambda_{11}+\lambda_{22}}{2}\pm\frac{1}{2}\sqrt{\left(\lambda_{11}-\lambda_{22}\right)^{2}+4\lambda_{12}\lambda_{21}+4\lambda_{13}\lambda_{31}} , \quad (23)$$

which remind the solutions for a two-band superconductor with renormalized (effective) interband coupling constants $\lambda_{12}\lambda_{21}\to\lambda_{12}\lambda_{21}+\lambda_{31}\lambda_{13}$. Such non-universal renormalization might explain the success of phenomenological two-band models. Here, *independent* of the signs of the interband couplings, $T_c$ is always *enhanced* by the third band coupled to one band, only.

# 4. The upper critical field of a three-band superconductor

Now we turn to the investigation of the most important magnetic property, i.e. the upper critical field $H_{c2}$. We assume that the vector potential $\boldsymbol{A}=\left(0,\boldsymbol{H}x,0\right)$, so the magnetic field is directed along the $z$ axis.

We will look for solutions of the GL equations (6) in the form $\Delta_i = C_i \exp\left(-\frac{x^2}{2\xi^2}\right)$.

This yields a system of linearized equations for the determination of $H_{c2}$:

$$\begin{cases} \left(\alpha_1 - \frac{2\pi K_1 H_{c2}}{\Phi_0}\right) C_1 + \gamma_{12} C_2 + \gamma_{13} C_3 = 0, \\ \gamma_{21} C_1 + \left(\alpha_2 - \frac{2\pi K_2 H_{c2}}{\Phi_0}\right) C_2 + \gamma_{23} C_3 = 0, \\ \gamma_{31} C_1 + \gamma_{32} C_2 + \left(\alpha_3 - \frac{2\pi K_3 H_{c2}}{\Phi_0}\right) C_3 = 0. \end{cases} \tag{24}$$

Introducing the dimensionless parameters $h_{c2} = \frac{2\pi K_1 H_{c2}}{\Phi_0}$, $d_{12} = \frac{D_2}{D_1}$, $d_{13} = \frac{D_3}{D_1}$, $n_{12} = \frac{N_2}{N_1}$, $n_{13} = \frac{N_3}{N_1}$, and $\frac{T}{T_c} = t$, we obtain: $H_{c2}$:

$$\begin{cases} \left(l - a_1 - h_{c2}\right) C_1 + \tilde{\gamma}_{12} n_{12} C_2 + \tilde{\gamma}_{13} n_{13} C_3 = 0, \\ \tilde{\gamma}_{21} C_1 + \left(l - a_2 - d_{12} h_{c2}\right) n_{12} C_2 + \tilde{\gamma}_{23} n_{13} C_3 = 0, \\ \tilde{\gamma}_{31} C_1 + \tilde{\gamma}_{32} n_{12} C_2 + \left(l - a_3 - d_{13} h_{c2}\right) n_{13} C_3 = 0. \end{cases} \tag{25}$$

Here $l = \ln\left(\frac{2\gamma \langle \omega_0 \rangle}{\pi T}\right) \approx 1 + \frac{1}{\Lambda^{(r)}} - t$. From Eq. (25) we obtain the general equation for $h_{c2}$:

$$\det \begin{pmatrix} 1 + \frac{1}{\Lambda^{(r)}} - t - a_1 - h_{c2} & \tilde{\gamma}_{12} n_{12} & \tilde{\gamma}_{13} n_{13} \\ \tilde{\gamma}_{21} & \left(1 + \frac{1}{\Lambda^{(r)}} - t - a_2 - d_{12} h_{c2}\right) n_{12} & \tilde{\gamma}_{23} n_{13} \\ \tilde{\gamma}_{31} & \tilde{\gamma}_{32} n_{12} & \left(1 + \frac{1}{\Lambda^{(r)}} - t - a_3 - d_{13} h_{c2}\right) n_{13} \end{pmatrix} = 0 \tag{26}$$

Next, we derive for each case which was considered above, the *T*-dependence of the upper critical field. To understand the influence of the third band on $h_{c2}(t)$, we consider the same dependence for a two-band superconductor with the matrix $\begin{pmatrix} \lambda_{11} & \lambda_{12} \\ \lambda_{21} & \lambda_{22} \end{pmatrix}$. Note that for a two-band superconductor $h_{c2}(t)$ doesn't depend on the sign of the interband interactions. It will be shown below that this degeneracy is *lifted* by the presence of a third band, at least for the cases we considered in section 3.

**(i)** For both cases we adopt $n_{12} = n_{13} = 1$ and $a_1 = a_2 = a_3 = a = \frac{k_0 + k_1}{k_0^2 + k_0 k_1 - 2k_1^2}$, $\tilde{\gamma}_{12} = \tilde{\gamma}_{21} = \tilde{\gamma}_{13} = \tilde{\gamma}_{31} = \tilde{\gamma}_{23} = \tilde{\gamma}_{32} = \tilde{\gamma} = \frac{k_1}{k_0^2 + k_0 k_1 - 2k_1^2}$ and $\eta_a = 1 + \frac{1}{k_0 + 2k_1}$ for attractive interband couplings and, $\eta_r = 1 + \frac{1}{k_0 + |k_1|}$ for repulsive ones. Taking into account these redefinitions we simplify Eq. (26):

$$A^{(h)} h_{c2}^3 + B^{(h)} h_{c2}^2 + C^{(h)} h_{c2} + D^{(h)} = 0\,, \tag{27}$$

$$A^{(h)} = -d_{12} d_{13}\,, \tag{28}$$

$$B^{(h)} = \left(d_{12} + d_{13} + d_{12} d_{13}\right)\left(\eta - a - t\right), \tag{29}$$

$$C^{(h)} = -\left(1 + d_{12} + d_{13}\right)\left(\eta - a - \tilde{\gamma} - t\right)\left(\eta - a + \tilde{\gamma} - t\right), \tag{30}$$

$$D^{(h)} = \left(\eta - a + 2\tilde{\gamma} - t\right)\left(\eta - a - \tilde{\gamma} - t\right)^2. \tag{31}$$

Based on the numerical solution of Eq. (27) we plot the *T*-dependencies of the upper critical field for the cases 1) with very small and very large ratios of diffusion coefficients for weak and strong interband coupling (see Fig. 2). We remind that for a two-band superconductor there is no difference between attractive and repulsive interband interactions.

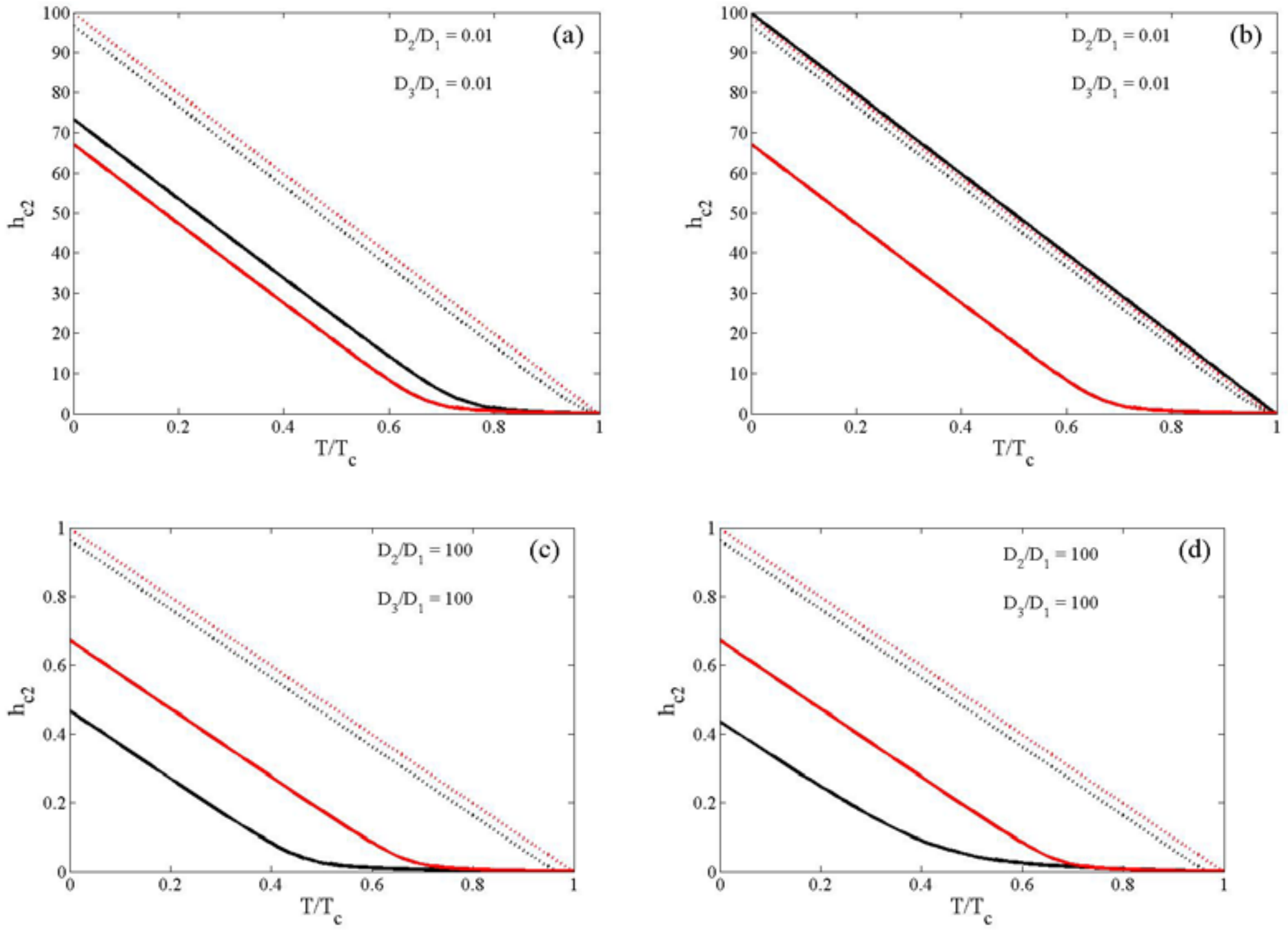

Fig. 2 (Color). *T*-dependencies of the upper critical field of a two-band (red lines) and three-band superconductor (black lines) with vanishing and very large ratios of diffusive coefficients for attractive (a, c) and repulsive (b, d) interaction. Solid lines are strong attractive interaction with $k_0 = 1$ and $k_1 = 0.3$ (panel a, c) and strong repulsive interactions with $k_0 = 1$ and $k_1 = -0.3$ (panel b, d). Dotted line are weak attractive and repulsive interaction with $k_0 = 1$ and $k_1 = 0.003$ (panel a, c) and $k_0 = 1$ and $k_1 = -0.003$ (panel b, d), respectively.

Strong repulsive interaction leads to an increase of $H_{c2}(0)$ for very small $d_{12}$ and $d_{13}$ while weak (attractive or repulsive) interband interactions don't change practically the magnitude of $H_{c2}(0)$ and the linear shape of $h_{c2}(t)$ (see Fig. 2 (a), (b)). Also we note that for the case of very large $d_{12}$ and $d_{13}$ there are almost no differences in the temperature evolution of $h_{c2}(t)$ between the repulsive and attractive interaction (see Fig. 2 (c), (d)). In addition, a strong interband coupling sufficiently decreases the slope of $h_{c2}(t)$ near $T_c$ for both values of $d_{12}$ and $d_{13}$ especially in the case of an attractive interaction. For a weak interaction, regardless of the sign of the interband coupling and the values of $d_{12}$ and $d_{13}$, we observe a linear dependence of $h_{c2}(t)$ with a constant slope, which slightly decreases in the close vicinity of $T_c$.

In order to clarify possible reasons for the very small slope of $h_{c2}(t)$ in the vicinity of $T_c$ for two- and three-band superconductors with strong interband interactions, we determine the temperature, at which the slope of $h_{c2}(t)$ starts to increase strongly. For this purpose we use the minimum curvature of $h_{c2}(t)$ in the interval $t \in [0,1]$, i.e. we solve the equation $\frac{d^3 h_{c2}}{dt^3} = 0$.

First we solve this equation for a two-band superconductor with the matrix of intra- and interband coefficients $\begin{pmatrix} k_0 & k_1 \\ k_1 & k_0 \end{pmatrix}$. The temperature $T_*^{(2)}$, where this sharp transition takes place is defined by the expression:

$$\frac{T_*^{(2)}}{T_c} = 1 - \frac{|k_1|}{k_0^2 - k_1^2}. \tag{32}$$

It's interesting to note that $T_*^{(2)}$ doesn't depend on the ratio of the diffusive constants for a two-band superconductor. From Eq. (32) we get for $k_0 = 1$ and $k_1 = 0.3$ (strong interband interaction) $T_*^{(2)} \approx 0.67 T_c$ in accordance with the data from Fig. 2 (solid red lines) and for $k_0 = 1$ and $k_1 = 0.003$ (weak interband interaction) we obtain $T_*^{(2)} \approx 0.997 T_c$ (dotted red lines), very close to $T_c$.

For a three-band superconductor the temperature, where the $h_{c2}(t)$ dependence shows a maximum value of the curvature, is determined by the expressions

$$\frac{T_*^{(3+)}}{T_c} = 1 - \frac{k_1(2d-1)}{(d-1)\left(k_0^2 + k_0 k_1 - 2k_1^2\right)}, \tag{33}$$

for attractive interband interactions ($k_1 > 0$) and

$$\frac{T_*^{(3-)}}{T_c} = 1 - \frac{|k_1|(d-2)}{(d-1)\left(k_0^2 + k_0 k_1 - 2k_1^2\right)}, \tag{34}$$

for repulsive ($k_1 < 0$) ones. Solving Eq. (27) to determine the third derivative, we assumed $d_{12} = d_{13} = d$ for the sake of simplicity. For instance, for $d = 0.01$, $k_0 = 1$ and $k_1 = 0.3$ (strong attractive interaction) from the expression (33) we get $T_*^{(3+)} \approx 0.73 T_c$ (solid black line in Fig. 2a) and for $k_0 = 1$ and $k_1 = 0.003$ (weak attractive interaction) $T_*^{(3+)} \approx 0.997T_c$ (dotted black line on the Fig. 2b); for $d = 100$ and the same sets of the attractive interaction constants we obtain $T_*^{(3+)} \approx 0.46T_c$ and $T_*^{(3+)} \approx 0.994T_c$ in accordance with the data from Fig. 2c. If $d = 0.01$, $k_0 = 1$ and $k_1 = -0.3$ (strong repulsive interaction) the expression (34) would give a negative value of $T_*^{(3-)}$, which means that there is no curvature in the investigated temperature interval (solid black line in Fig. 2b), while for $k_0 = 1$ and $k_1 = -0.003$ (weak repulsive interaction) we obtain $T_*^{(3-)} \approx 0.994T_c$ (dotted black line in Fig. 2b). Finally, for $d = 100$ $k_0 = 1$ and $k_1 = -0.3$ we have $T_*^{(3-)} \approx 0.43T_c$ and for $k_0 = 1$ and $k_1 = -0.003$ with the same $d$ the minimum curvature of $h_{c2}$ appears at $T_*^{(3-)} \approx 0.997T_c$.

In Appendix C it is shown that for any set of the interband coupling constants $\frac{d^2 h_{c2}(t)}{dt^2} \geq 0$, which means that for a three-band superconductor with the matrix of interaction constants $\begin{pmatrix} k_0 & k_1 & k_1 \\ k_1 & k_0 & k_1 \\ k_1 & k_1 & k_0 \end{pmatrix}$, there are no inflection points on the $h_{c2}(t)$-curves (at least within the GL approach) and the upper critical field shows always an upward curvature in sharp contrast with a single-band superconductor.

**(ii)** From the matrix of interaction coefficients we have $a_1 = -\frac{k_2}{2k_1^2}$, $a_2 = a_3 = -\frac{1}{2k_2}$, $\tilde{\gamma}_{12} = \tilde{\gamma}_{21} = \tilde{\gamma}_{13} = \tilde{\gamma}_{31} = -\frac{1}{2k_1}$, $\tilde{\gamma}_{23} = \tilde{\gamma}_{32} = -\frac{1}{2k_2}$. Here we applied numerical solution of Eq. (26) and plotted $h_{c2}(t)$ for limiting cases of vanishing and very large $d_{12}$ and $d_{13}$ (see Fig. 3).

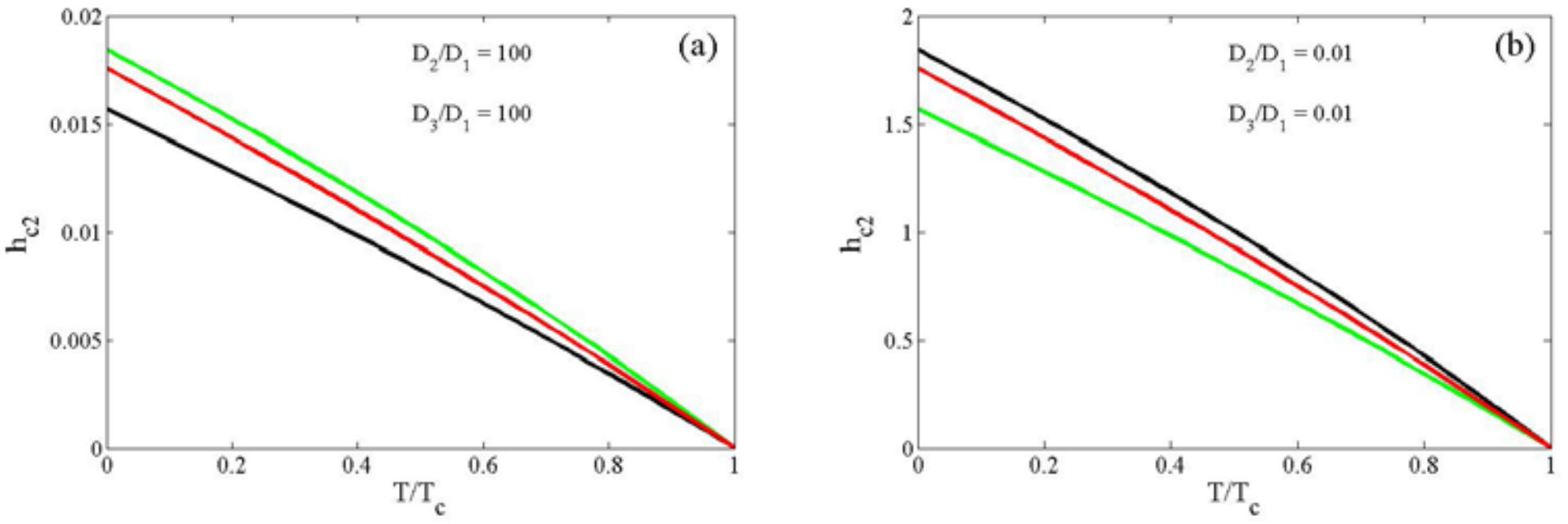

Fig. 3 (Color). *T*-dependencies of the upper critical field of a three-band superconductor without intraband pairing interactions and with attractive and repulsive interband interaction (black and green lines, respectively) for very large (panel a) and very small (panel b) ratios of diffusive coefficients (as shown in the legends). The interaction constants are $k_1 = 0.3$, $k_2 = 0.1$ and $k_1 = -0.3$, $k_2 = -0.1$ as compared with two-band superconductor (red lines).

Note that for such a matrix of the interaction coefficients slightly nonlinear dependencies with a negative curvature of the $h_{c2}(t)$ curves do occur at variance with the above considered cases. It's important to note that the temperature dependences of the upper critical field $h_{c2}^{(3+)}(t)$ and $h_{c2}^{(3-)}(t)$ for three-band superconductors with attractive and repulsive interband interaction split asymmetrically from the $h_{c2}^{(2)}(t)$ curve for the two-band superconductor (see Fig. 4).

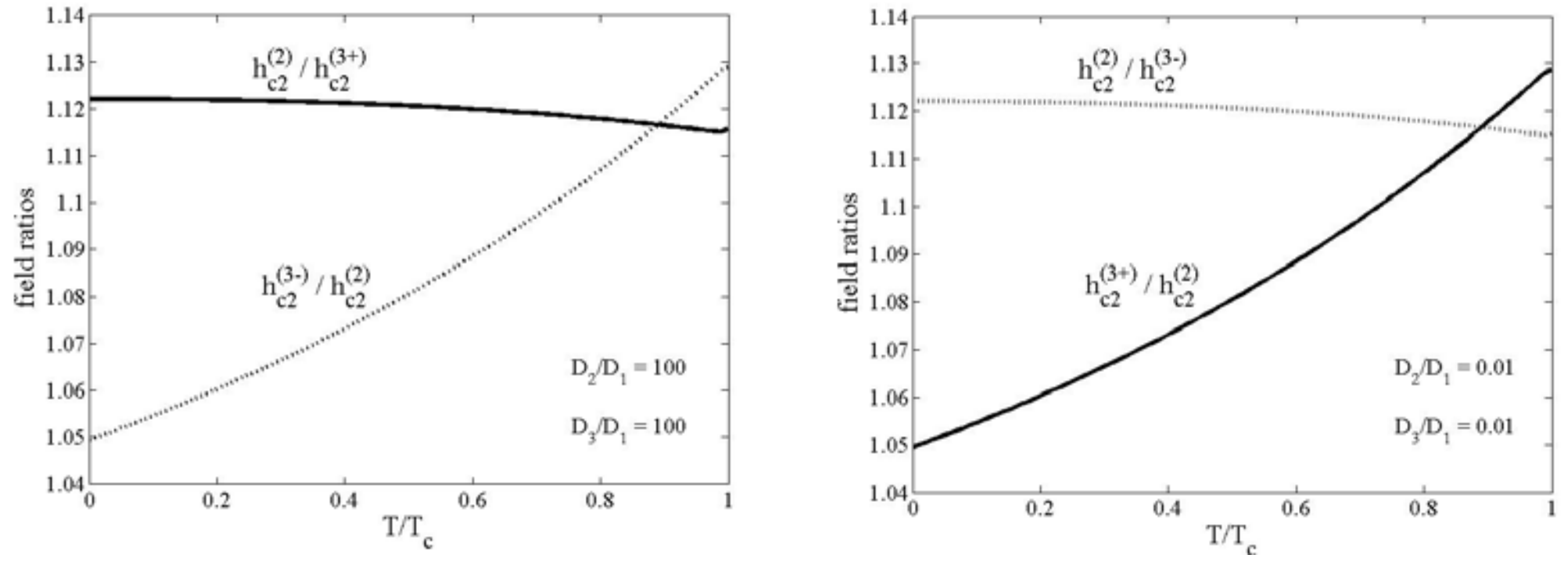

Fig. 4. The *T*-dependent ratios of upper critical fields for two-band and three-band superconductors in the case of different dirtyness of bands (as shown in the legends). Solid curves: three-band superconductor with attractive interband interactions. Dotted lines: the same as before with repulsive couplings.

Noteworthy, the ratio of $h_{c2}(t)$ for a two-band and three-band superconductor with attractive interband coupling for very large $d_{12}$ and $d_{13}$ do absolutely coincide with that for the inverse values (small) $d_{12}$ and $d_{13}$ for a three-band superconductor with repulsive interband interactions. Analogously, the ratio of $h_{c2}(t)$ for a three-band and two-band superconductor with repulsive interband couplings and very large $d_{12}$ and $d_{13}$ absolutely coincide also with the same ratio for very small $d_{12}$ and $d_{13}$ of a three-band superconductor with attractive interband coupling.

**(iii)** For the pseudo-three-band model we apply again the numerical solution of Eq. (26) for strong and weak attractive/repulsive interactions in the limits of vanishing and very large $d_{12}$ and $d_{13}$. In the numerical analysis we found out that there are no differences between repulsive and attractive interaction for all values of $\lambda_{ij}$ and the relationships for this model coincide with those for two-band superconductors on a qualitative level.

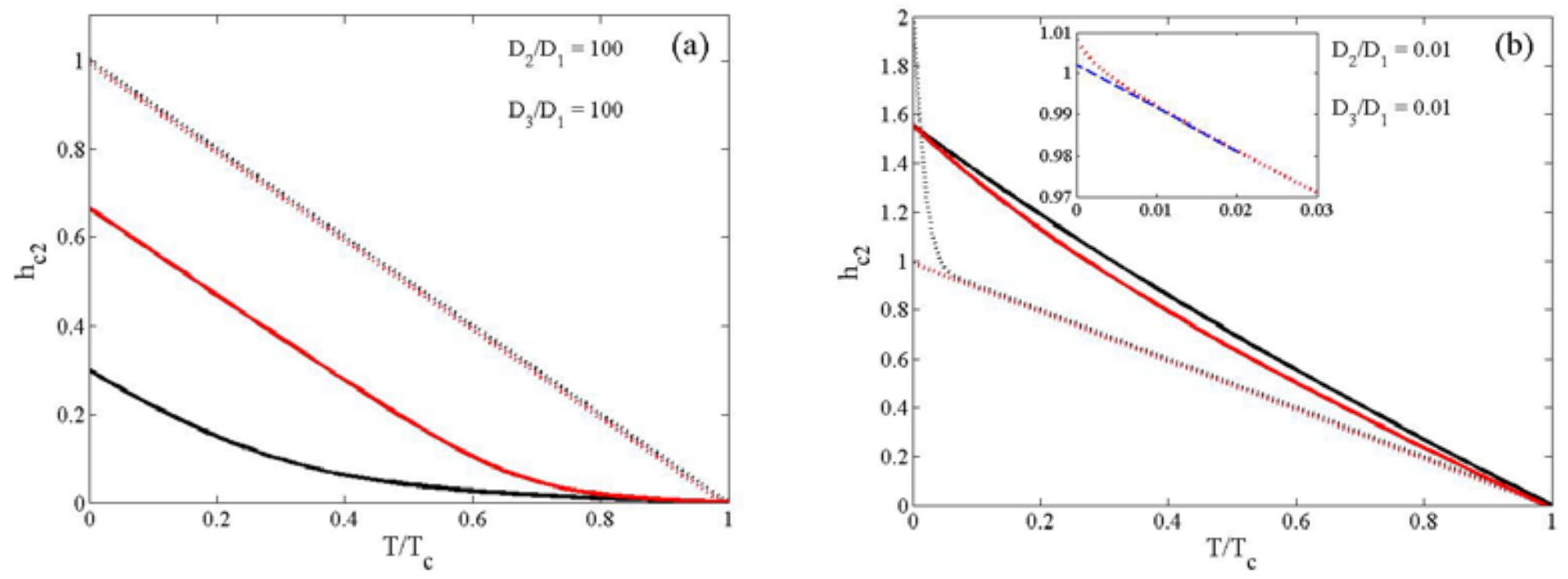

Fig. 5 (Color). *T*-dependencies of the upper critical field of a pseudo-three-band and a two-band superconductor (black and red lines, respectively) with very large (a) and vanishing (b) ratios of the diffusive coefficients (see legends). The coupling constants are $\lambda_{11} = 1$, $\lambda_{22} = \lambda_{33} = 0.5$ and $\lambda_{12} = \lambda_{21} = \lambda_{13} = \lambda_{31} = 0.3$ (solid lines) and $\lambda_{12} = \lambda_{21} = \lambda_{13} = \lambda_{31} = 0.003$ (dotted lines). The curves for the case of repulsive interband interactions fully coincide with the attractive counter part shown here (see text). Inset on the Fig. 6b represents the temperature behavior of the upper critical field at low temperature.

Another noteworthy feature of the obtained $h_{c2}(t)$ solutions shown in Fig.5 (b) is the strong increase of $h_{c2}(t)$ at low temperature for a three-band superconductor with weak interband coupling and small $d_{12}$ and $d_{13}$ resulting in a doubling of $H_{c2}(0)$ compared with $H_{c2}(0)$ of a two-band superconductor with the same parameters. It is also interesting to note that a very slight upturn of $h_{c2}(t)$ is already visible for the corresponding two-band superconductor at very low temperatures (see the inset in Fig. 5 (b)).

Furthermore, it is also seen that by a strong interband coupling $H_{c2}(0)$ is *enhanced* at low $d_{12}$ and $d_{13}$ but *reduced* at high $d_{12}$ and $d_{13}$. The latter effect is found to be more pronounced for three-band than for two-band superconductors.

# 5. Conclusions

Based on a microscopic formalism we have derived the Ginzburg-Landau equations for dirty intraband three-band superconductors. Within this approach we have examined the influence of a third band on the critical temperature and the temperature dependence of the upper critical field. We have considered some special cases of the matrix of interaction constants and have demonstrated explicitly the richness of three-band models as compared to frequently used two-band cases. In particularly, we have shown that in contrast to two-band superconductors the character (sign) of the interband interaction affects the value of the critical temperature and the temperature dependences of the upper critical field. The results of our analysis can be helpful for a characterization of the magnetic properties of some iron-based superconductors.

In future we plan to compare our results obtained within the three-band approach and the recently extended GL two-band formalism [34-36].

## Acknowledgement

We thank the German-Ukrainian project for financial support. Discussions with D. Efremov, B. Holzapfel, Jeroen van den Brink, and A. Omelyanchouk are kindly acknowledged. Y.Y. thanks the IFW-Dresden for hospitality where the present work has been performed.

# APPENDIX A: DERIVATION OF GINZBURG-LANDAU EQUATIONS

For the anomalous Green functions $f_i$ we get:

$$f_i = \frac{D_i}{2\omega^2}\nabla^2\Delta_i - \frac{|\Delta_i|^2\Delta_i}{2\omega^3} + \frac{\Delta_i}{\omega}. \qquad \text{(A1)}$$

Substituting $f_i$ into the self-consistency equations (4) after the summation over the Matsubara frequencies we have finally

$$\begin{cases} \Delta_1 N_1 = \lambda_{11} N_1 \left[ \frac{\pi D_1}{8T_c} \Pi^2 \Delta_1 - \frac{7\zeta(3)}{8\pi^2 T_c^2} |\Delta_1|^2 \Delta_1 + l\Delta_1 \right] + \lambda_{12} N_2 \left[ \frac{\pi D_2}{8T_c} \Pi^2 \Delta_2 - \frac{7\zeta(3)}{8\pi^2 T_c^2} |\Delta_2|^2 \Delta_2 + l\Delta_2 \right] + \\ \lambda_{13} N_3 \left[ \frac{\pi D_3}{8T_c} \Pi^2 \Delta_3 - \frac{7\zeta(3)}{8\pi^2 T_c^2} |\Delta_3|^2 \Delta_3 + l\Delta_3 \right], \\ \Delta_2 N_2 = \lambda_{21} N_1 \left[ \frac{\pi D_1}{8T_c} \Pi^2 \Delta_1 - \frac{7\zeta(3)}{8\pi^2 T_c^2} |\Delta_1|^2 \Delta_1 + l\Delta_1 \right] + \lambda_{22} N_2 \left[ \frac{\pi D_2}{8T_c} \Pi^2 \Delta_2 - \frac{7\zeta(3)}{8\pi^2 T_c^2} |\Delta_2|^2 \Delta_2 + l\Delta_2 \right] + \\ \lambda_{23} N_3 \left[ \frac{\pi D_3}{8T_c} \Pi^2 \Delta_3 - \frac{7\zeta(3)}{8\pi^2 T_c^2} |\Delta_3|^2 \Delta_3 + l\Delta_3 \right], \\ \Delta_3 N_3 = \lambda_{31} N_1 \left[ \frac{\pi D_1}{8T_c} \Pi^2 \Delta_1 - \frac{7\zeta(3)}{8\pi^2 T_c^2} |\Delta_1|^2 \Delta_1 + l\Delta_1 \right] + \lambda_{32} N_2 \left[ \frac{\pi D_2}{8T_c} \Pi^2 \Delta_2 - \frac{7\zeta(3)}{8\pi^2 T_c^2} |\Delta_2|^2 \Delta_2 + l\Delta_2 \right] + \\ \lambda_{33} N_3 \left[ \frac{\pi D_3}{8T_c} \Pi^2 \Delta_3 - \frac{7\zeta(3)}{8\pi^2 T_c^2} |\Delta_3|^2 \Delta_3 + l\Delta_3 \right]. \end{cases}$$

(A2)

In order to obtain the GL equations, we multiply the first equation by $\lambda_{22}\lambda_{33} - \lambda_{23}\lambda_{32}$ , the second equation by $-(\lambda_{12}\lambda_{33} - \lambda_{13}\lambda_{32})$ and the third equation by $\lambda_{12}\lambda_{23} - \lambda_{13}\lambda_{22}$. Then we sum these three expressions over Matsubara frequencies and finally obtain the GL equations (6) for $\Delta_1$, $\Delta_2$, and $\Delta_3$.

# APPENDIX B: THE DETERMINATION OF THE CRITICAL TEMPERATURE

$T_c$ can be found from the linearization of the GL system (6):

$$\begin{cases} \alpha_1 \Delta_1 + \gamma_{12} \Delta_2 + \gamma_{13} \Delta_3 = 0, \\ \alpha_2 \Delta_2 + \gamma_{21} \Delta_1 + \gamma_{23} \Delta_3 = 0, \\ \alpha_3 \Delta_3 + \gamma_{31} \Delta_1 + \gamma_{32} \Delta_2 = 0. \end{cases} \tag{B1}$$

which leads to the cubic equation:

$$\alpha_1\alpha_2\alpha_3 - \gamma_{32}\gamma_{23}\alpha_1 - \gamma_{31}\gamma_{13}\alpha_2 - \gamma_{12}\gamma_{21}\alpha_3 + \gamma_{12}\gamma_{31}\gamma_{23} + \gamma_{13}\gamma_{21}\gamma_{32} = 0, \tag{B2}$$

or using the phenomenological constants $\gamma_{ij}$ introduced in the main paper:

$$\left(l-a_1\right)\left(l-a_2\right)\left(l-a_3\right)-\tilde{\gamma}_{32}\tilde{\gamma}_{23}\left(l-a_1\right)-\tilde{\gamma}_{31}\tilde{\gamma}_{13}\left(l-a_2\right)-\tilde{\gamma}_{12}\tilde{\gamma}_{21}\left(l-a_3\right)+\tilde{\gamma}_{12}\tilde{\gamma}_{31}\tilde{\gamma}_{23}+\tilde{\gamma}_{13}\tilde{\gamma}_{21}\tilde{\gamma}_{32}=0 . \qquad \text{(B3)}$$

Eq. (B3) can be rewritten as:

$$\begin{aligned}&l^3-\left(a_1+a_2+a_3\right)l^2+\left(a_1a_2+a_1a_3+a_2a_3-\tilde{\gamma}_{12}\tilde{\gamma}_{21}-\tilde{\gamma}_{31}\tilde{\gamma}_{13}-\tilde{\gamma}_{32}\tilde{\gamma}_{23}\right)l+\\&\tilde{\gamma}_{32}\tilde{\gamma}_{23}a_1+\tilde{\gamma}_{31}\tilde{\gamma}_{13}a_2+\tilde{\gamma}_{12}\tilde{\gamma}_{21}a_3+\tilde{\gamma}_{12}\tilde{\gamma}_{31}\tilde{\gamma}_{23}+\tilde{\gamma}_{13}\tilde{\gamma}_{21}\tilde{\gamma}_{32}=0.\end{aligned} \qquad \text{(B4)}$$

Comparing Eq. (B3) with the general form of a cubic equation and bearing in mind the representation of the coefficients $a_i$ and $\gamma_{ij}$, we get

$$l^3+Bl^2+Cl+D=0, \qquad \text{(B5)}$$

where

$$B\equiv-\frac{1}{\det\left(\lambda\right)}\sum_i M_{ii} ,$$

$$C\equiv\frac{M_{11}M_{22}+M_{11}M_{33}+M_{22}M_{33}-M_{12}M_{21}-M_{13}M_{31}-M_{23}M_{32}}{\det^2\left(\lambda\right)} ,$$

$$D\equiv\frac{M_{11}M_{23}M_{32}+M_{13}M_{22}M_{31}+M_{12}M_{21}M_{33}-M_{13}M_{21}M_{32}-M_{12}M_{23}M_{31}-M_{11}M_{22}M_{33}}{\det^3\left(\lambda\right)} .$$

Depending on the sign of the discriminant $Z=18BCD-4B^3D+B^2C^2-4C^3-27D^2$ we have three distinct real roots, if $Z>0$, one real root and two complex conjugate roots, if $Z<0$ and three real roots, if $Z=0$.

If we expand the coefficients $B$, $C$ and $D$ in terms of the coupling constants $\lambda_{ij}$ and simplify the obtained expressions, we reveal that Eq. (B5) for determination of the critical temperature is equivalent to the secular equation for the coupling matrix (7).

# APPENDIX C: THE CURVATURE OF THE UPPER CRITICAL FIELD

In the case (**i**) the upper critical field is determined by the equation:

$$A^{(h)}h_{c2}^3 + B^{(h)}h_{c2}^2 + C^{(h)}h_{c2} + D^{(h)} = 0\,, \tag{C1}$$

where

$$A^{(h)} = -d_{12}d_{13}\,, \tag{C2}$$

$$B^{(h)} = \left(d_{12} + d_{13} + d_{12}d_{13}\right)\left(\eta - a - t\right), \tag{C3}$$

$$C^{(h)} = -\left(1 + d_{12} + d_{13}\right)\left(\eta - a - \tilde{\gamma} - t\right)\left(\eta - a + \tilde{\gamma} - t\right), \tag{C4}$$

$$D^{(h)} = \left(\eta - a + 2\tilde{\gamma} - t\right)\left(\eta - a - \tilde{\gamma} - t\right)^2. \tag{C5}$$

Let's consider the simple case when $d_{12} = d_{13} = d$ and $a = \dfrac{k_0 + k_1}{k_0^2 + k_0k_1 - 2k_1^2}$, $\tilde{\gamma} = \dfrac{k_1}{k_0^2 + k_0k_1 - 2k_1^2}$ and $\eta_a = 1 + \dfrac{1}{k_0 + 2k_1}$ (attractive interband interaction).

Then the second derivative yields

$$\frac{d^2h_{c2}}{dt^2} = \frac{1}{2}\frac{1}{d\left(k_0^2 + k_0k_1 - 2k_1^2\right)}\times$$

$$\frac{8d\left(d-1\right)^2 k_1^2\left(k_0^2 + k_0k_1 - 2k_1^2\right)^2}{\left(\left(t - \dfrac{\left(d-1\right)\left(k_0^2 + k_0k_1 - 2k_1^2\right) - 2dk_1 + k_1 - 2\sqrt{2d}\,ik_1}{\left(k_0^2 + k_0k_1 - 2k_1^2\right)\left(d-1\right)}\right)\left(t - \dfrac{\left(d-1\right)\left(k_0^2 + k_0k_1 - 2k_1^2\right) - 2dk_1 + k_1 + 2\sqrt{2d}\,ik_1}{\left(k_0^2 + k_0k_1 - 2k_1^2\right)\left(d-1\right)}\right)\right)^{\frac{3}{2}}}$$

(C6)

After further simplifications we obtain

$$\frac{d^2 h_{c2}}{dt^2} = \frac{1}{2}\frac{1}{d\left(k_0^2 + k_0 k_1 - 2k_1^2\right)}\frac{8d\left(d-1\right)^2 k_1^2 \left(k_0^2 + k_0 k_1 - 2k_1^2\right)^5}{\left(\left(\left(k_0^2 + k_0 k_1 - 2k_1^2\right)\left(d-1\right)^2 t - \left(d-1\right)\left(k_0^2 + k_0 k_1 - 2k_1^2\right) - 2dk_1 + k_1\right)^2 + 8dk_1^2\right)^{\frac{3}{2}}} =$$

$$\frac{4\left(d-1\right)^2 k_1^2 \left(k_0^2 + k_0 k_1 - 2k_1^2\right)^4}{\left(\left(\left(k_0^2 + k_0 k_1 - 2k_1^2\right)\left(d-1\right) t - \left(d-1\right)\left(k_0^2 + k_0 k_1 - 2k_1^2\right) - 2dk_1 + k_1\right)^2 + 8dk_1^2\right)^{\frac{3}{2}}} \geq 0.$$

(C7)

Analogously it can be shown for repulsive interband interactions, that $\frac{d^2 h_{c2}}{dt^2}$ is non-negative, too.

# REFERENCES

1. Y. Kamihara, T. Watanabe, M. Hirano, and H. Hosono, J. Am. Chem. Soc. **130**, 3296 (2008).

2. M. Tortello, D. Daghero, G. A. Ummarino, V. A. Stepanov, J. Jiang, J. D. Weiss, E. E. Hellstrom, and R. S. Gonnelli, Phys. Rev. Lett. **105**, 237002 (2010).

3. M. Abdel-Hafiez, L. Harnagea, V. Grinenko, V.B. Zabolotnyy, D. Bombor, Y. Krupskaya, C. Hess, S. Wurmehl, A.U.B. Wolter, B. Büchner, S.-L. Drechsler, H. Rosner, and S. Johnston to be published.

4. P. Cudazzo, G. Profeta, A. Sanna, A. Floris, A. Continenza, S. Massidda, and E. K. U. Gross, Phys. Rev. Lett. **100**, 257001 (2008).

5. M. J. Rice, Phil. Mag. Part B **65**, 1419 (1992).

6. M. J. Rice, H.-Y. Choi, and Y.R. Wang, Phys. Rev. B **44**, 10414 (1991).

7. Q.H. Wang, C. Platt, Y. Yang, F.C. Zhang, W. Hanke, T.M. Rice, and R. Thomale, arXiv:1305.2317v1.

8. J. Garaud, J. Carlström and E. Babaev, Phys. Rev. Lett. **10**7, 197001 (2011).

9. Shi-Zeng Lin and Xiao Hu, New J. Phys. **14**, 063021 (2012).

10. J. Garaud, J. Carlström, E. Babaev, and M. Speight, Phys. Rev. B **87**, 014507 (2013).

11. L. Komendová, Yajiang Chen, A. A. Shanenko, M. V. Milošević, F. M. Peeters, Phys. Rev. Lett. **108**, 207002 (2012).

12. S. V. Shulga and S.-L. Drechsler, J. of Low. Temp. Phys. **129**, 93 (2002).

13. S.V. Shulga, S.-L. Drechsler, G. Fuchs, K.-H. Müller, K. Winzer, M. Heinecke, K. Krug, Phys. Rev. Lett. **80**, 1730 (1998).

14. H. Doh, M. Sigrist, B.K. Cho, and S.L. Lee, Phys. Rev. Lett. **83**, 5350 (1999).

15. I.N. Askerzade, A. Gencer, N. Guclu, and A. Kihc, Supercond. Sci. Technol. **15**, L13 (2002).

16. A. Gurevich, Phys. Rev. B **67**, 184515 (2003).

17. A.A. Golubov, A.E. Koshelev, Phys. Rev. B **68**, 104503 (2003).

18. M. Mansor, J.P. Carbotte, Phys. Rev. B **72**, 024538 (2005).

19. I.N. Askerzade, JETP Lett. **81**, 583 (2005).

20. P. Udomsamuthirun, A. Changjan, C. Kumvongsa, S. Yoksan, Physica C **433**, 62 (2006).

21. A. Gurevich, Physica C **456**, 160 (2007).
22. L. Min-Zia, G. ZiZhao, Chin. Phys. **16**, 826 (2007).
23. A. Changjana, P. Udomsamuthiruna, Solid State Commun. **151**, 988 (2011).
24. Yanagisawa Y. Tanaka, I. Hase, and K. Yamaji, J. Phys. Soc. Jpn. **81**, 024712 (2012).
25. R.G. Dias and A.M. Marques, Supercond. Sci. Technol. **24**, 085009 (2011).
26. V. Stanev and Z. Tešanović, Phys. Rev. B **81**, 134522 (2010).
27. V. Stanev, Phys. Rev. B **85**, 174520 (2012).
28. S.-Z. Lin and X. Hu, Phys. Rev. Lett. **108**, 177005 (2012).
29. M. Nitta, M. Eto, T. Fujimori, K. Ohashi, J. Phys. Soc. Jpn. **81**, 084711 (2012).
30. T.K. Ng, Phys. Rev. Lett. 103, 236402 (2009).
31. V. Stanev and A.E. Koshelev, arXiv:1306.4268v1.
32. K.D. Usadel, Phys. Rev. Lett. **25**, 507 (1970).
33. A.V. Svidzinski, "Space-inhomogeneous problems in the theory of superconductivity", Nauka, Moscow (1982).
34. A. A. Shanenko, M. V. Milošević, F. M. Peeters, and A. V. Vagov, Phys. Rev. Lett. **106**, 047005 (2011).
35. A. Vagov, A. A. Shanenko, M. V. Milošević, V. M. Axt, and F. M. Peeters, Phys. Rev. B **86**, 144514 (2012).
36. N. V. Orlova, A. A. Shanenko, M. V. Milošević, and F. M. Peeters, A. V. Vagov and V. M. Axt, Phys. Rev. B **87**, 134510 (2013).